\documentstyle[aasms4]{article}

\slugcomment{Accepted for publication in \it The Astrophysical Journal, Letters}

\lefthead{Gibson, Maloney \& Sakai}
\righthead{Blending and the Cepheid Distance Scale}

\begin{document}

\def\kms{km\,s$^{-1}$}
\def\spose#1{\hbox to 0pt{#1\hss}}
\def\simlt{\mathrel{\spose{\lower 3pt\hbox{$\mathchar"218$}}
     \raise 2.0pt\hbox{$\mathchar"13C$}}}
\def\simgt{\mathrel{\spose{\lower 3pt\hbox{$\mathchar"218$}}
	  \raise 2.0pt\hbox{$\mathchar"13E$}}}
\def\eg{{\rm e.g.}}
\def\ie{{\rm i.e.}}
\def\etal{{\rm et~al.}}

\title{Has Blending Compromised Cepheid-Based Determinations of 
the Extragalactic Distance Scale?}

\author{Brad K. Gibson and Philip R. Maloney}
\affil{Center for Astrophysics \& Space Astronomy, 
University of Colorado,
Campus Box 389, Boulder, Colorado, 80309-0389}
\and
\author{Shoko Sakai}
\affil{Kitt Peak National Observatory, National Optical Astronomy Observatories,
Tucson, Arizona, 85726}

\begin{abstract}
We examine the suggestion that half of the \it HST Key Project\rm- and
\it Sandage/Saha\rm-observed galaxies have had their distances
systematically underestimated, by $0.1-0.3$\,mag in the distance
modulus, due to the underappreciated influence of stellar profile
blending on the WFC chips. The signature of such an effect would be a
systematic trend in (i) the Type Ia supernovae corrected peak
luminosity and (ii) the Tully-Fisher residuals, with increasing
calibrator distance, and (iii) a differential offset between PC and
WFC distance moduli, within the same galaxy. The absence of a trend
would be expected if blending were negligible (as has been inherently
assumed in the analyses of the aforementioned teams). We adopt a
functional form for the predicted influence of blending that is
consistent with the models of Mochejska \etal\ and Stanek \& Udalski,
and demonstrate that the expected correlation with distance predicted
by these studies is not supported by the data. We conclude that the
Cepheid-based extragalactic distance scale has \it not \rm been
severely compromised by the neglect of blending.
\end{abstract}


\section{Introduction}
\label{introduction}

In recent papers, Mochejska \etal\ (1999) and Stanek \& Udalski (1999)
suggest that blending of stellar images on \it HST \rm WFC images
has led the \it HST Key Project on the Extragalactic Distance Scale
\rm (Gibson \etal\ 2000; Sakai \etal\ 2000) and \it Type Ia Supernovae
Calibration Team \rm (Saha \etal\ 1999; hereafter \it Sandage/Saha\rm)
to systematically underestimate the distance to the galaxies in their
sample, resulting in an overestimate of the Hubble Constant. The
magnitude of the predicted effect increases with galaxy distance,
ranging from 0.05\,mag for the nearest galaxies to 0.35\,mag for the
most distant galaxies in the samples (\ie, distance modulus
$\mu_\circ\approx 32$). To date, blending has been assumed to be a
negligible contributor to the Cepheid distance scale systematic error
budget. The results of Mochejska \etal\ and Stanek \& Udalski,
however, suggest that both Hubble Constant teams overestimated
$H_\circ$ by $5-10$\%. Since the magnitude of this suggested effect is
as large as the entire quoted standard error budget (\eg, Table~6 of
Sakai \etal\ 2000), it is crucial that further empirical tests of the
Mochejska \etal\ and Stanek \& Udalski blending scenario be made.

In Section \ref{analysis}, we describe a simple, yet heretofore
neglected, empirical test of the Mochejska \etal\ (1999) and Stanek \&
Udalski (1999) blending scenario. By examining the distribution of
$V$-band corrected peak luminosities for the 8 Type Ia supernovae
(SNe) calibrators used in the $H_\circ$ analysis by Gibson \etal\
(2000), the distribution of $I$- and $H$-band Tully-Fisher residuals,
for the 18 \it HST\rm-observed calibrators used by Sakai \etal\
(2000), and the comparison of Planetary Camera (PC) and Wide Field
Camera (WFC) distance moduli for five suitable Virgo and Fornax
Cluster galaxies, we demonstrate that the systematic trend predicted
by Mochejska \etal\ and Stanek \& Udalski is \it not \rm supported by
the data. The absence of a trend is consistent with the inherent
assumptions of both \it HST \rm Hubble Constant teams. Section
\ref{summary} summarizes our findings.

\section{Analysis}
\label{analysis}

We quantify the predicted influence that blending has upon
Cepheid-based distance determinations by employing the models of
Stanek \& Udalski (1999, Figure~2); their ``$>$5\%'' cutoff models
(\ie, a minimum of $5\%$ of the mean flux of a Cepheid is required for
a star to be included as a blend) are suitable for our needs. The
blending difference $\Delta\mu_\circ$ between what Stanek \& Udalski
label the ``true'' distance, and that ``measured'' by the \it HST Key
Project \rm and \it Sandage/Saha Team \rm is represented by the
following functional form:
\begin{equation}
\Delta\mu_\circ\equiv\mu_\circ({\rm true})-\mu_\circ({\rm
measured})\approx
0.002\,d^{1.62} \quad{\rm mag}, \label{eq:eq1}
\end{equation}
\noindent
where $d$ is the galaxy distance in Mpc -- \ie, the ``measured''
distance modulus is hypothesized to systematically underestimate the
true modulus. The magnitude of this effect is 0.1\,mag at
$d=12$\,Mpc, and 0.3\,mag at $d=23$\,Mpc.

A direct consequence of blending of the magnitude predicted by Stanek
\& Udalski (1999) and Mochejska \etal\ (1999) (for the remainder of
this paper we shall refer to these papers collectively as Stanek
\etal) is that a systematic trend in the distribution of both
corrected Type Ia SNe peak luminosities -- assumed to be standard
candles -- and Tully-Fisher residuals, as a function of calibrator
distance, should result. Empirically, we should observe the corrected
peak luminosities and the T-F residuals shift to an apparently lower
luminosity with increasing calibrator distance. This test was not
included in the Mochejska \etal\ or Stanek \& Udalski analyses.

In Figure~\ref{fig:fig1}, the corrected $V$-band peak luminosities of
the eight Type Ia SNe calibrators (Gibson \etal\ 2000) are shown as a
function of the ``measured'' $\mu_\circ$; the three circled points
refer to those SNe with poorer-quality light curves and/or
photometry. We have calculated least-square fits with and without
their inclusion. The solid horizontal curve -- the weighted mean of
all eight points -- represents the locus about
which symmetric scatter is expected, in the absence of significant
blending effects. The dotted curve shows the Stanek \etal\ blending
prediction, given by equation \ref{eq:eq1}; if blending is important,
this dotted curve should be the locus about which the calibrators
scatter symmetrically. A straight-line fit, with a slope statistically
indistinguishable from zero (as assumed in the previous distance-scale
analyses), provides an excellent representation of the data. (In fact,
the reduced chi-square values $\chi_\nu$ for the straight-line fits
are so low, $\sim 0.2-0.4$, that one would ordinarily conclude that
the error bars on the data points have been overestimated. Monte Carlo
simulations of the data set suggest this overestimate is a factor of
roughly $1.4-2$.)

\placefigure{fig:fig1}

If the zero-point of the blending model curve is taken to be the
corrected $V$-band peak luminosity zero-point defined by the ``no
blending'' value, then the blending prediction is inconsistent with
the data at the $2.2\sigma$ level (this is true for any zero-point
fainter than M$_V^{\rm corr,ZP}$$\approx$-19.5), using either the five
high-quality calibrators or the full set of eight. However, if
blending is important, there is no {\it a priori} justification for
assuming that the zero point of the blending model curve in
Figure~\ref{fig:fig1} is equal to the ``no-blending'' value. If we
relax this assumption and allow M$_V^{\rm corr,ZP}$ to vary,
$\chi_\nu$ for the Stanek \etal\ blending model reaches an acceptable
value for a zero point M$_V^{\rm corr,ZP}=-19.62\pm 0.1$. The dotted
curve of Figure~\ref{fig:fig1} shows this best-fit blending solution.
While this latter fit is not
formally as significant as the ``no blending'' straight line, it is
excluded only at the $\sim1\sigma$ level\footnote{If the error bars
have been overestimated, the range of M$_V^{\rm corr,ZP}$ for which
acceptable solutions can be found for the blending model shrinks and
the quality of the fit is lower; however, the blending model is ruled
out at the $2\sigma$ level, regardless of zero-point, only if the
overestimate is as large as a factor of two.}.

In allowing M$_V^{\rm corr,ZP}$ to vary freely, however, we are
ignoring another constraint imposed by the observations, namely, the
observed tight clustering of the five high-quality data points about a
zero-slope line with M$_V^{\rm corr,ZP}=-19.46$. We have therefore
carried out Monte Carlo simulations of the data set, assuming that the
blending model is correct, with the distribution of points in
$\mu_\circ$ chosen to match the data. The dispersion of the points
about the blending model curve is chosen to be either equal to the
typical error bar, or that value reduced by a factor of 1.4, in case
the errors have been overestimated. We generated a large number of
synthetic data sets of five supernovae, and then asked in what
fraction of the data sets at least four out of the five fall as
closely to a zero-slope line (with any zero-point) as the actual
data. This fraction is about $5\%$ using the actual errors and about
$12\%$ for the reduced dispersion. However, unsurprisingly, the
probability that the data points will scatter about a zero-slope line
with an intercept close to $-19.46$ drops rapidly as we move the zero
point away from this value. For all values of M$_V^{\rm corr,ZP}$ for
which a reasonable fit of the blending model is possible, the
probability of reproducing this aspect of the data (\ie, of obtaining
a zero-slope line with an intercept within $0.1$ of $-19.46$) is less
than a few percent (for either choice of dispersion), so that we again
rule out the blending model at the $\sim 2\sigma$ level.

In Figure \ref{fig:fig2}, we show the distribution of $I$- (left
panel) and $H$-band (right panel) Tully-Fisher residuals with
$\mu_\circ$, for the 18 \it HST\rm-observed calibrators used by Sakai
\etal\ (2000). The three ground-based calibrators are not included in
the formal blending analysis, since equation~\ref{eq:eq1} holds only
for blended \it HST \rm WFC frames, and not these particular
ground-based datasets. The solid and dotted curves in
Figure~\ref{fig:fig2} have the same meaning as those shown in
Figure~\ref{fig:fig1}; although the blending curve is shown with an
intercept of zero, this is not necessarily the case, and the curve has
been included only to show the shape of the blending prediction. In
calculating least-square fits to the data points, we have assumed an
intrinsic dispersion of $0.2$\,mag in the $I$- and $H$-band
Tully-Fisher relations (Sakai \etal\ 2000).

\placefigure{fig:fig2}

As with the ground-based calibrators, we have not included NGC~4603
(in the Cen30 cluster) in our analysis, despite Newman et~al.'s (1999)
recent Cepheid distance determination of $\mu_\circ=32.61\pm
0.11$\,mag. In combination with the $H$-band photometry and 21cm
line-width measurements tabulated by Han (1992), the Newman
et~al. modulus implies an $H$-band Tully-Fisher residual of
$\sim$1.3$\pm$0.3\,mag - \ie, NGC~4603 is \it extremely \rm discrepant
with the $H$-band Tully-Fisher relation presented in Sakai \etal\
(2000, equation~10). This discrepancy {\it is} in the direction
expected for data compromised by severe blending.

Two crucial points need to be mentioned here, however: first, the
Stanek \etal\ functional form for blending (equation \ref{eq:eq1}) was
based upon Cepheids on the WFC chips only; in contrast, the Newman
\etal\ analysis included only Cepheid candidates found on the
higher-resolution PC chip. The functional form for PC Cepheid
blending would naturally lie between the two curves shown in
Figure~\ref{fig:fig2} - \ie, the implied $H$-band residual for
NGC~4603 is highly discrepant with both the no-blending assumption \it
and \rm the Stanek \etal\ scenario, \it assuming the Newman
et~al. modulus is correct\rm. Second, and more importantly, NGC~4603 is
\it not \rm employed as an $H_\circ$ calibrator, so whether or not
its Cepheid distance has been underestimated is irrelevant in terms of
the $H_\circ$ analyses of the \it HST Key Project \rm and \it
Sandage/Saha\rm. We speculate that the Newman \etal\ result has indeed
been compromised by extreme crowding effects, such that its distance
has been underestimated by $\simgt$1\,mag. This would require an
intrinsic stellar background contamination several times greater than
the LMC and M31 fields employed by Stanek \etal\ in deriving
equation~\ref{eq:eq1}, which themselves are already several times
greater than the typical \it HST Key Project \rm field. To reiterate,
this has {\it no} impact upon $H_\circ$, as NGC~4603 has not been used
as a calibrator.

The inferior statistics associated with the Tully-Fisher residuals (in
both bands) do not allow us to make any unequivocal statements
regarding the Stanek \etal\ blending scenario.\footnote{Ferrarese
\etal\ (2000; Appendix~A) present an analog to our
Figure~\ref{fig:fig2}, but conclude that the Stanek \etal\ blending
hypothesis can be excluded at the 1.85\,$\sigma$ level, in apparent
contradiction with our analysis. The source of the discrepancy can be
traced to the fact that Ferrarese \etal\ neglect the uncertainties
associated with each of the $H$-band Tully-Fisher residuals shown in
their Figure~18 - \ie, they performed a least squares fit to the
twenty points shown, ignoring the plotted error bars.  Doing so does
indeed return a slope in the opposite direction predicted by Stanek
et~al., but the conclusion to be drawn by such an approach is not that
Stanek \etal\ can be ruled out at the 1.85\,$\sigma$ level, but that
the assumption of ``no blending'' (as assumed by the \it Key Project
\rm and \it Sandage/Saha\rm) can be ruled out at the 1.85\,$\sigma$
level, and that Stanek \etal\ can be ruled out at some unspecified
amount greater than 1.85\,$\sigma$.  The significance of this
conclusion, though, is rendered moot by the neglect of the
uncertainties associated with the individual residuals in the
Ferrarese et~al. analysis.} While the best-fitting straight lines
actually have the opposite slope to the blending prediction, both the
blending model and a zero-slope line differ from the best-fit lines at
most at the $1\sigma$ level. More data at larger values of
$\mu_\circ$, where there is a larger difference between the blending
and no-blending models, are needed. At present all we can state is
that the Tully-Fisher residuals do not provide any support for the
claim that the distance scale has been systematically underestimated
due to the effects of stellar blending.

Independent of these \it direct \rm tests of the Stanek \etal\
blending scenario, there exists a more straightforward \it
differential \rm test. Because of the factor of two improvement in
spatial resolution between the Planetary Camera (0.046\,arcsec/pixel)
and the Wide Field Camera (0.1\,arcsec/pixel), if Stanek \etal\ are
correct, a systematic difference should be seen between the distance moduli
based upon PC Cepheids alone, $\mu_\circ$(PC), and those based upon WFC
Cepheids alone, $\mu_\circ$(WF). The sense of the systematic offsest
should be such that $\mu_\circ$(PC)$>$$\mu_\circ$(WF). For galaxies
in the Virgo or Fornax Clusters, under the Stanek \etal\ blending
hypothesis, the measured $\mu_\circ$(WF) should underestimate the true
value by 0.16\,mag, while the measured $\mu_\circ$(PC) would be
underestimated by only 0.05\,mag - \ie, there should be a 0.11\,mag
differential offset between the measured $\mu_\circ$(PC) and
$\mu_\circ$(WF). This prediction is independent of any arguments
pertaining to Tully-Fisher residuals or Type Ia SNe corrected peak
luminosities (Figures~\ref{fig:fig1} and \ref{fig:fig2}).

An examination of the full \it HST Key Project \rm and \it
Sandage/Saha \rm samples showed that only five galaxies met the
criteria necessary to perform the above differential test. First, we
required the galaxy be distant enough ($\mu_\circ>30.7$) to show at
least a 0.1\,mag differential effect between $\mu_\circ$(PC) and
$\mu_\circ$(WF) and, second, we insisted there be at least four
high-quality Cepheids on the PC, in order to define a useful
period-luminosity relation.\footnote{We employed the same quality,
color, and period cuts employed by the original authors, when drawing
up the Cepheid sample of Table~\ref{tbl:tbl1}.} Table~\ref{tbl:tbl1}
lists the five galaxies (four in Virgo and one in Fornax) employed in
our differential test (NGC~1365, 4536, 4496A, 4321, and 4548).
Recall, the predicted magnitude of $\mu_\circ$(PC)$-$$\mu_\circ$(WF),
for Virgo/Fornax galaxies, is $+$0.11\,mag. As can be seen in
Table~\ref{tbl:tbl1}, though, the weighted mean offset between
$\mu_\circ$(PC) and $\mu_\circ$(WF), for these five galaxies, is
$+$0.002$\pm$0.049. In other words, this differential test allows us
to reject the Stanek \etal\ blending scenario at the 2.2\,$\sigma$
level. The significance of this conclusion can be improved in the
future with the addition of further calibrators in the $\mu_\circ=31$
regime, provided reasonable numbers of PC Cepheids can be uncovered
(\ie, distant $\mu_\circ>32$ galaxies are not required).

\placetable{tbl:tbl1}

\section{Summary}
\label{summary}

We have undertaken a simple, empirical, test of the suggestion of
Mochejska \etal\ (1999) and Stanek \& Udalski (1999) that blending has
seriously compromised the extragalactic Cepheid distance scale as
measured by the \it HST Key Project on the Extragalactic Distance
Scale \rm and the \it Sandage/Saha Type Ia SNe Calibration Team\rm.
Mochejska \etal, Stanek \& Udalski, and Paczynski (1999) have each
speculated that the magnitude of this systematic effect could be as
large as $\sim$0.2\,mag (\ie, $\sim$10\%), meaning that both Hubble
Constant teams have overestimated $H_\circ$ by the same amount.

The distributions of both the Tully-Fisher residuals and $V$-band Type
Ia SNe corrected peak luminosities, for their respective nearby
calibrators, show no discernible dependence upon distance. This
behavior is consistent with both Hubble Constant teams' inherent
assumption that blending plays a negligible role in the global error
budget, and does not support the Mochejska \etal\ (1999) and Stanek \&
Udalski (1999) blending scenario. Formally, the Type Ia SNe corrected
peak luminosities allow us to reject the Stanek \etal\ blending
scenario at the 2.2\,$\sigma$ level, assuming a corrected $V$-band
peak luminosity zero point $\simgt$$-$19.5. If the SNe zero point is
0.16\,mag brighter than currently assumed, statistically acceptable
fits for the blending model are possible. However, the probability of
reproducing the small scatter of the data about a zero-slope line with
M$_V^{\rm corr,ZP}=-19.46$ (the weighted mean value) is sufficiently
low in this latter case that we can still rule out the model at the
$\sim 2\sigma$ level. We present evidence that the uncertainties in
the SNe peak luminosity may be overestimated by as much as a factor of
two; if so, the blending hypothesis can be ruled out at the $2\sigma$
level regardless of SNe zero point. The absence of a systematic offset
between Cepheid distance moduli derived from the higher resolution
Planetary Camera and the Wide Field Camera for galaxies with suitably
large $\mu_\circ$ also rules out the blending model at the $2.2\sigma$
level.

It seems likely that the discrepancy between the Stanek \etal\
prediction and the data, as we have discussed in this paper, is due to
the high stellar background associated with the LMC and M31 fields
used in their analyses; these background levels are \it not \rm
representative of the more distant \it HST \rm WFC frames. While it
is almost certainly true that \it some \rm fields/galaxies have been
compromised by blending effects, this does \it not \rm appear to be a
global phenomemon which has compromised the Cepheid-based distance
scale.

We note in passing that our \it empirical \rm test results are
supported on \it theoretical \rm grounds by the artificial star tests
described by Ferrarese \etal\ (2000) - these tests suggest that the
blending bias is $\sim$0.02\,mag, even in the most crowded \it HST \rm
WFC fields.

\acknowledgements{BKG acknowledges support from the NASA Long-Term
Space Astrophysics Program (NAG5-7262) and the FUSE Science Team
(NAS5-32985). PRM is supported by the NASA Astrophysical Theory
Program under grant NAG5-4061 and by NSF under grant AST-9900871. SS
acknowledges support from NASA through the Long-Term Space
Astrophysics Program NAS-7-1260}

\newpage

\begin{figure}
\caption{Distribution of $V$-band peak luminosities (corrected for
light curve shape) for the 8 Type Ia SNe calibrators used by Gibson
\etal\ (2000). The uncertainties reflect those associated with the SNe
photometry, light curve shape, line-of-sight reddening and host-galaxy
period-luminosity dispersion, as defined by entry `l' in Table~7 of
Gibson \etal\ (2000). The three calibrators with sub-standard
photometry or light curve quality (SN~1972E, 1960F, and 1974G) are
denoted with large open circles. The solid horizontal line represents
the weighted mean of all 8 calibrators (corresponding to M$_V^{\rm
corr,ZP}=-19.46$), and is the expected locus if blending is
negligible, as has been implicitly assumed by both the \it HST Key
Project on the Extragalactic Distance Scale \rm and the \it
Sandage/Saha Team\rm.  The dotted curve represents the predicted
behavior under the ``Stanek \etal'' (\ie, Mochejska \etal\ 1999 and
Stanek \& Udalski 1999) blending hypothesis, with an adopted zero
point of M$_V^{\rm corr,ZP}=-19.62$; this latter zero point yields an
acceptable statistical fit to the data, albeit at a significance level
lower than that favored by the ``no blending'' hypothesis.}
\label{fig:fig1}
\end{figure}

\begin{figure}
\caption{Distribution of $I$- (left panel) and $H$-band (right panel)
Tully-Fisher residuals for the 18 \it HST\rm-observed calibrators used
by Sakai \etal\ (2000). The uncertainties reflect those of the
photometry and line width for the calibrator in question (from Table~2 of
Sakai \etal), as well as that of the intrinsic dispersion of the
$I$- and $H$-band Tully-Fisher relations (also from Sakai \etal).
The solid horizontal line represents the
expected locus about which scattering should occur if blending is
negligible, as has been implicitly assumed by both the \it HST Key
Project on the Extragalactic Distance Scale \rm and the \it
Sandage/Saha Team\rm.  The dotted curve represents the predicted
behavior under the ``Stanek \etal'' (\ie, Mochejska \etal\ 1999 and
Stanek \& Udalski 1999) blending hypothesis; the $y-$intercept for this
curve is not necessarily zero. Formally the data are
completely consistent with a line of zero slope, but the blending
model is discrepant at worst at the $1\sigma$ level.}
\label{fig:fig2}
\end{figure}

\clearpage

\begin{deluxetable}{lccc}
\tablecaption{Differential Test of Blending at the Distance of the Virgo
Cluster \label{tbl:tbl1}}
\tablewidth{0pt}
\tablehead{
\colhead{Galaxy} & 
\colhead{$\mu_\circ$(PC)\tablenotemark{a}} &
\colhead{$\mu_\circ$(WF)\tablenotemark{a}} &
\colhead{$\mu_\circ$(PC)$-$$\mu_\circ$(WF)}
}
\startdata
NGC~1365  & 31.307$\pm$0.127 (9) & 31.305$\pm$0.090 (25) & $+$0.002$\pm$0.156\\
NGC~4536  & 30.851$\pm$0.097 (6) & 30.886$\pm$0.046 (21) & $-$0.035$\pm$0.167\\
NGC~4496A & 30.980$\pm$0.104 (6) & 30.949$\pm$0.043 (45) & $+$0.031$\pm$0.113\\
NGC~4321  & 31.154$\pm$0.181 (6) & 30.960$\pm$0.070 (35) & $+$0.194$\pm$0.194\\
NGC~4548  & 30.990$\pm$0.034 (4) & 31.009$\pm$0.065 (20) & $-$0.019$\pm$0.073\\\\\
&  \multicolumn{2}{r}{Weighted Mean}  & $+$0.002$\pm$0.049\\
&  \multicolumn{2}{r}{Stanek et~al. Blending Prediction} & $+$0.11$\qquad\quad\;\;$\\
\enddata
\tablenotetext{a}{The number of Cepheids employed in the period-luminosity
fitting is noted in the parentheses adjacent to the true distance modulus.}
\end{deluxetable}

\clearpage

\epsscale{1.0}
\plotone{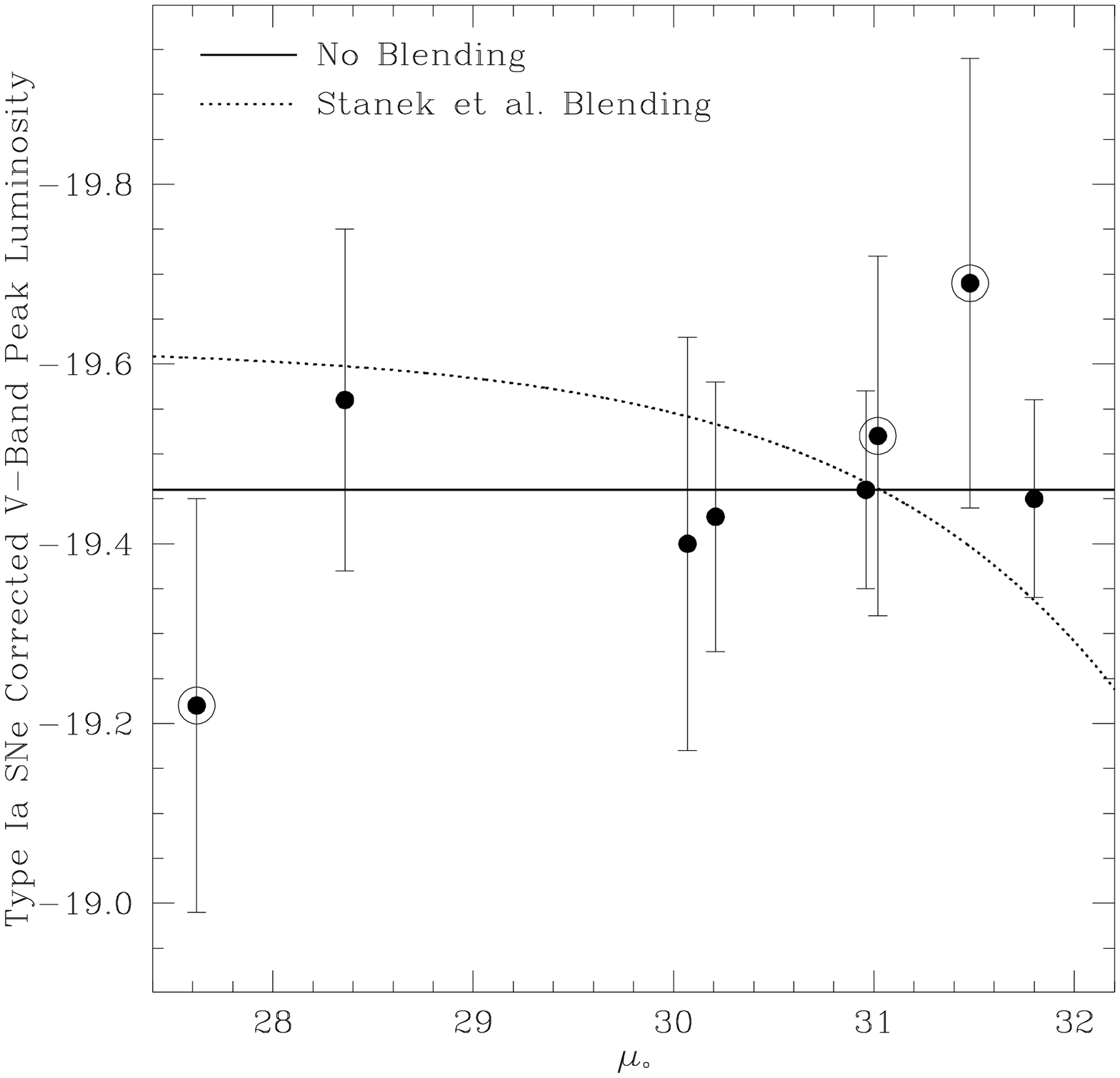}

\clearpage

\epsscale{1.0}
\plotone{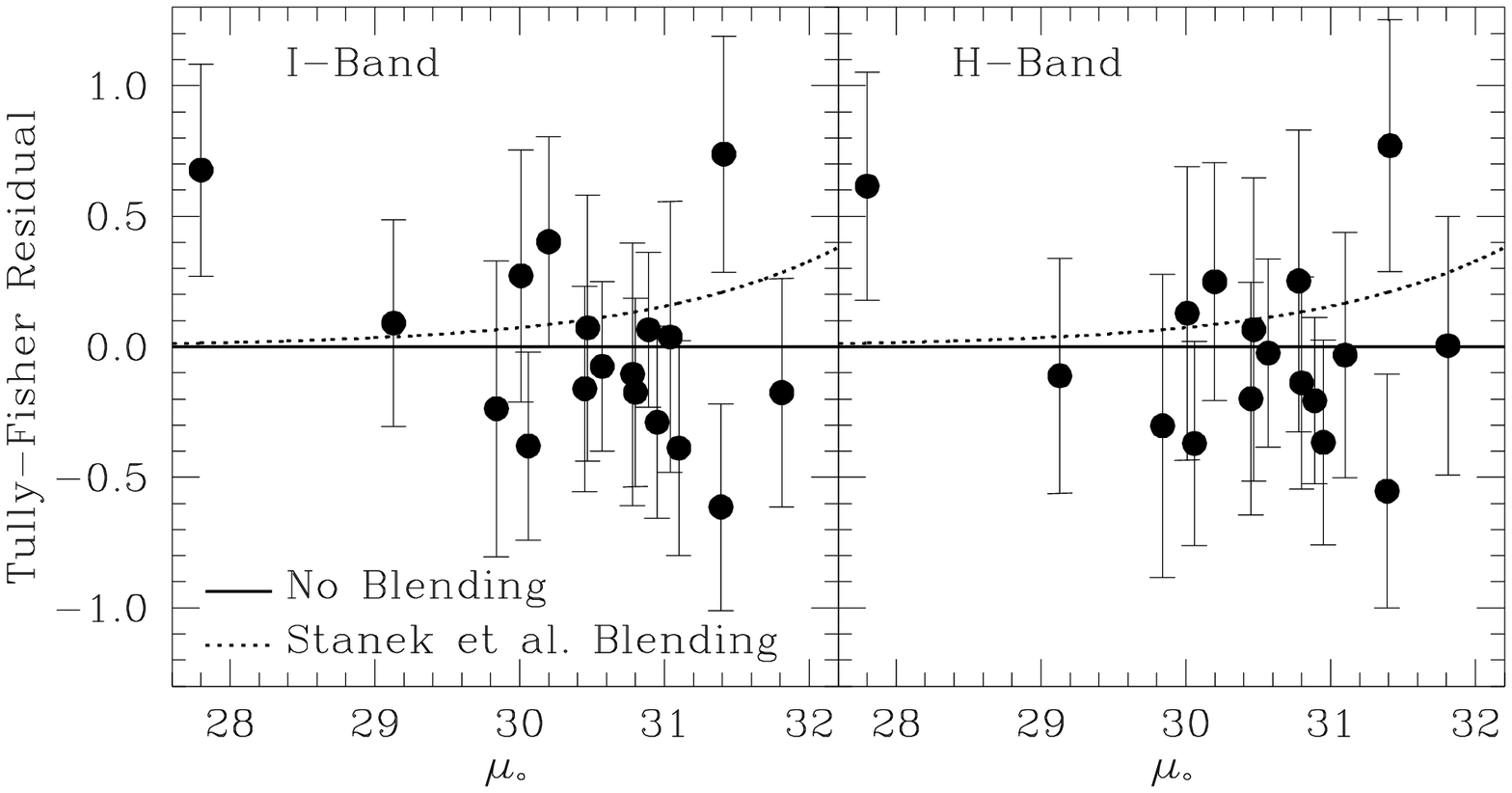}


\newpage

\begin{thebibliography}{}

\bibitem[laura]{F00} 
Ferrarese, L., et~al. 2000, \pasp, in press (astro-ph/9911193)

\bibitem[brad]{G00}
Gibson, B.K., et~al. 2000, \apj, in press (astro-ph/9908149)

\bibitem[han]{H92}
Han, M. 1992, \apjs, 81, 35

\bibitem[someone]{MMSS99}
Mochejska, B.J., Macri, L.M., Sasselov, D.D. \& Stanek, K.Z. 1999, 
preprint (astro-ph/9908293)

\bibitem[newman]{New99}
Newman, J.A., Zepf, S., Davis, M., Freedman, W.L., Madore, B.F., Stetson, P.B.,
Silbermann, N. \& Phelps, R. 1999, \apj, 523, 506

\bibitem[bohdan]{P99}
Pacynski, B. 1999, \nat, 401, 331

\bibitem[abi]{S99} 
Saha, A., Sandage, A., Tammann, G.A., Labhardt, L., Macchetto, F.D. \& 
Panagia, N. 1999, \apj, 522, 802

\bibitem[shoko]{S00} 
Sakai, S., et~al. 2000, \apj, in press (astro-ph/9909269)

\bibitem[kris]{SU99}
Stanek, K.Z. \& Udalski, A. 1999, preprint (astro-ph/9909346)

\end{thebibliography}
\end{document}